\documentclass[aps,pra,twocolumn,superscriptaddress,longbibliography,showpacs]{revtex4-1}
\usepackage{amsmath,amssymb,wasysym,graphicx}
\usepackage[utf8]{inputenc}
\usepackage[T1]{fontenc}
\usepackage{xcolor}

\IfFileExists{newtxtext.sty}
{\usepackage{newtxtext,newtxmath}}
{\IfFileExists{stix.sty}
	{\usepackage{stix}}
	{\IfFileExists{mathptmx.sty}
		{\usepackage{mathptmx}}{} } }

\usepackage{textcomp}

\usepackage{bm}

\IfFileExists{siunitx.sty}{\usepackage{booktabs,siunitx}}{}

\usepackage{color}
\definecolor{LinkColor}{rgb}{0.256,0.439,0.588}
\usepackage{hyperref}
\hypersetup{
	pdfauthor={good guys},
	pdftitle={good title},
	colorlinks=true,
	citecolor=LinkColor,
	linkcolor=LinkColor,
	urlcolor=LinkColor
}

\newcommand{\beq} {\begin{equation}}
\newcommand{\eeq} {\end{equation}}
\newcommand{\bea} {\begin{eqnarray}}
\newcommand{\eea} {\end{eqnarray}}
\newcommand{\be} {\begin{equation}}
\newcommand{\ee} {\end{equation}}

\begin{document}
\title{Quantum criticality of a $\mathbb{Z}_{3}$ symmetric spin chain with long-range interactions}

\author{Xue-Jia Yu}
\affiliation{International Center for Quantum Materials, School of Physics, Peking University, Beijing 100871, China}
\author{Chengxiang Ding}
\affiliation{School of Science and Engineering of Mathematics and Physics, Anhui University of Technology, Maanshan, Anhui 243002, China}
\author{Limei Xu}
\email{limei.xu@pku.edu.cn}
\affiliation{International Center for Quantum Materials, School of Physics, Peking University, Beijing 100871, China}
\affiliation{Collaborative Innovation Center of Quantum Matter, Beijing, China}
\affiliation{Interdisciplinary Institute of Light-Element Quantum Materials and Research Center for Light-Element Advanced Materials, Peking University, Beijing, China}  
\date{\today}

\begin{abstract}
 Based on large-scale density matrix renormalization group techniques, we investigate the critical behaviors of quantum three-state Potts chains with long-range interactions. Using fidelity susceptibility as an indicator, we obtain a complete phase diagram of the system. The results show that as the long-range interaction power $\alpha$ increases, the critical points $f_{c}^{*}$ shift towards lower values. In addition, the critical threshold $\alpha_{c}(\approx 1.43$) of the long-range interaction power is obtained for the first time by a non-perturbative numerical method. This indicates that the critical behavior of the system can be naturally divided into two distinct universality classes, namely the long-range ($\alpha \textless \alpha_c$) and short-range ($\alpha \textgreater \alpha_c$) universality classes, qualitatively consistent with the classical $\phi^{3}$ effective field theory. This work provides a useful reference for further research on phase transitions in quantum spin chains with long-range interaction.
\end{abstract}

\maketitle

\section{INTRODUCTION}%
\label{sec:introduction}
 Quantum phase transitions (QPTs) are phase transitions between quantum matters at zero temperature by tuning athermal parameters, which can be a first-order phase transition represented by some sudden abrupt jump behavior or a continuous phase transition described by a critical exponent. Universality class categorized by critical points or unstable fixed points in the sense of renormalization group (RG)~\cite{cardy1996scaling} is a core concept in QPTs. Using field theory or numerical exact approaches, conventional or unconventional QPT can be described by constructing simplified effective lattice models~\cite{xu2012unconventional,yu2022emergent,guo2022pra}. Therefore, quantum many-body systems with nearest-neighbor interactions, such as the transverse field Ising model, Heisenberg model, and Hubbard model, are of fundamental importance for understanding QPTs and universality classes~\cite{sachdev_2011}. A well-known QPT is the second-order Ising transition in the one-dimensional transverse field Ising model, and its critical exponents are perfectly supported by experimental results~\cite{sachdev_2011}.  


Quantum systems with long-range interactions, such as Coulomb interaction ($1/r_{ij}$)~\cite{saffman2010rmp}, dipole-dipole interaction ($1/r_{ij}^3$)~\cite{deng2005pra,Lahaye_2009}, and van der Waals interaction ($1/r_{ij}^6$)~\cite{saffman2010rmp}, have attracted widespread attention in recent years, accompanied by significant advances in experimental techniques for manipulating quantum simulators, such as atomic, molecular and optical systems~\cite{rydberg_exp,Lahaye_2009,ritschrmp2013,Carr_2009,blatt2012quantum}. 
For instance, tunable power-law interactions $1/r^{d+\alpha}$ with a power $0\le \alpha+d \le 3$ are realized in trapped ions~\cite{britton2012engineered,islam2013emergence,richerme2014non,jurcevic2014quasiparticle,song_arxiv2301}, which provides a perfect platform for studying novel physics of quantum many-body systems with long-range power-law interaction and stimulated many subsequent many-body physics studies. One example is the neutral Rydberg atom trapped in optical tweezers with programmable van der Waals interactions. It provides promising tunable platforms to explore various novel physics, such as gapped $\mathbb{Z}_{2}$ quantum spin liquids~\cite{ruben2021prx,Samajdar2022,semeghini2021probing,samajdar2021quantum,Slagle2022,cheng2021}, quantum phase transitions between different density wave ordered (e.g. $\mathbb{Z}_{3}$ ordered) and disordered phases~\cite{Samajdar2019PRL,Samajdae2018pra,Seth2018prb,Marcin2021,Matthew2021,roger2021,Slagle2021prb}.
Specifically, the $\mathbb{Z}_{n}$ symmetric quantum spin model system  ~\cite{alicea2016topological,Fendley_2012}, namely the "parafermion" model system, favors topological phases with more efficient non-Abelian anyon bound states~\cite{Fendley_2012,alicea2016topological},  providing a possible approach for universal topological quantum computing, thereby attracting extensive attention and stimulated extensive studies~\cite{Fendley_2012,notA2020,Phasediagramz32015prb,fendleyscipost2020,Yu_cbc_2021,Vaeziprx2014,weiliprb2015}. However, despite extensive interest in $\mathbb{Z}_{n}$ symmetric quantum many-body systems with long-range interactions, it remains challenging to fully understand their critical behavior both theoretically and numerically.



 It is well known that a $d$ dimensional quantum system with short-range interactions has a well-known equivalent classical counterpart in $d + 1$ dimensions. However, the quantum system with long-range interactions does not have a direct counterpart due to the subtle relationship between classical and quantum critical behaviors. For classical O(N) or $\mathbb{Z}_{n}$ symmetric spin model systems with long-range interactions ~\cite{Fisher1972prl,Knap2013prl,Defenu2017prb,Theumann1985prb,defenu2109,defenu2015pre,brezin2014crossover,angelini2014pre,behan2017prl,Behan_2017}, previous RG calculations show that according to the interaction power $\alpha$, the
 critical behavior falls into three university classes 
  namely, 1) the mean-field universality class when $\alpha \le d/2$, 2) the long-range universality class when $d/2 \textless \alpha \le \alpha_c$, and 3) the short-range universality class for $\alpha \textgreater \alpha_c $. Note that region 2) is a 'non-classical' region where the critical behavior is characterized by a peculiar long-range critical exponent
  $\alpha_c(=2-\eta_{SR})$, which can be predicted by perturbative RG calculations with a short-range anomalous exponent $\eta_{SR}$  ~\cite{Fisher1972prl,defenu2109}.
 The quantum three-state Potts chain is the simplest example of "parafermion" systems, which shows a continuous phase transition from 'topological phase' (Potts ordered) to trivial phase (disordered),  
 thereby is of crucial important for quantum computing~\cite{Fendley_2012,alicea2016topological,francesco2012conformal,Ginsparg_CFT}. The question is what is the critical behavior of such quantum Potts chains with long-range interaction in the "non-classical" region, and how to estimate the critical exponent $\alpha_{c}$ if there is a long-range to short-range universality class crossover.

Fidelity susceptibility is a purely geometric quantity of quantum states from quantum information world with an obvious advantage that no prior knowledge of order parameters and symmetry-breaking is required. It has been applied to detect a wide range of QPTs~\cite{Yu2014,Albuquerque2010prb,Schwandt2009prl,Yu2009pre,sun2015prb,onig2016prb,Wei2019pra,Lv2022pra,yu2022_rydberg_prb,sun2019prb,Tu2022,sun2022biorthogonal} induced by a sudden change in the structure of the wave-function. The fidelity susceptibility, defined as the response of the wavefunction overlap of two neighboring ground states with respect to an external field, diverges at the critical point and is almost zero away from the critical point, thus characterizing the QPTs well. 
For example, 
experiments detect the QPTs in terms of fidelity susceptibility by using the neutron scattering or angle-resolved photoemission spectroscopy (ARPES) techniques~\cite{Gu_2014}. Here, we investigate the finite-size scaling behavior of the fidelity susceptibility~\cite{gu2010fidelity,Gu2009pre,Gu_2009,You_2015} in the quantum Potts chain with long-range interactions using the finite-size density-matrix renormalization-group (DMRG) method~\cite{White1992prl,Schollw2005rmp,SCHOLLWOCK201196} based on the matrix product states (MPS) ~\cite{SCHOLLWOCK201196,Verstraete2004prl}.  The critical long-range interaction power $\alpha_c$ is determined in a non-perturbative way for the first time, providing important insight into phase transitions of quantum spin chains with long-range interaction.



The rest of this paper is organized as follows:in Sec.~\ref{sec:model} contains the lattice model of quantum Potts chain with long-range power-law interaction, the numerical method employed, and the scaling relations of fidelity susceptibility. Sec.~\ref{sec:phase} shows the phase diagram of the quantum Potts chain with long-range interaction and the finite-size scaling of the critical behavior, followed by a brief discussion in comparison with previous two-loop RG results. The conclusion is presented in Sec.~\ref{sec:summary}. Additional data for our numerical calculations are provided in the Appendixes.

\section{MODEL AND METHODS}%
\label{sec:model}

\subsection{Quantum Potts chain with long-range interaction}

\begin{figure}[tb]
\includegraphics[width=0.48\textwidth]{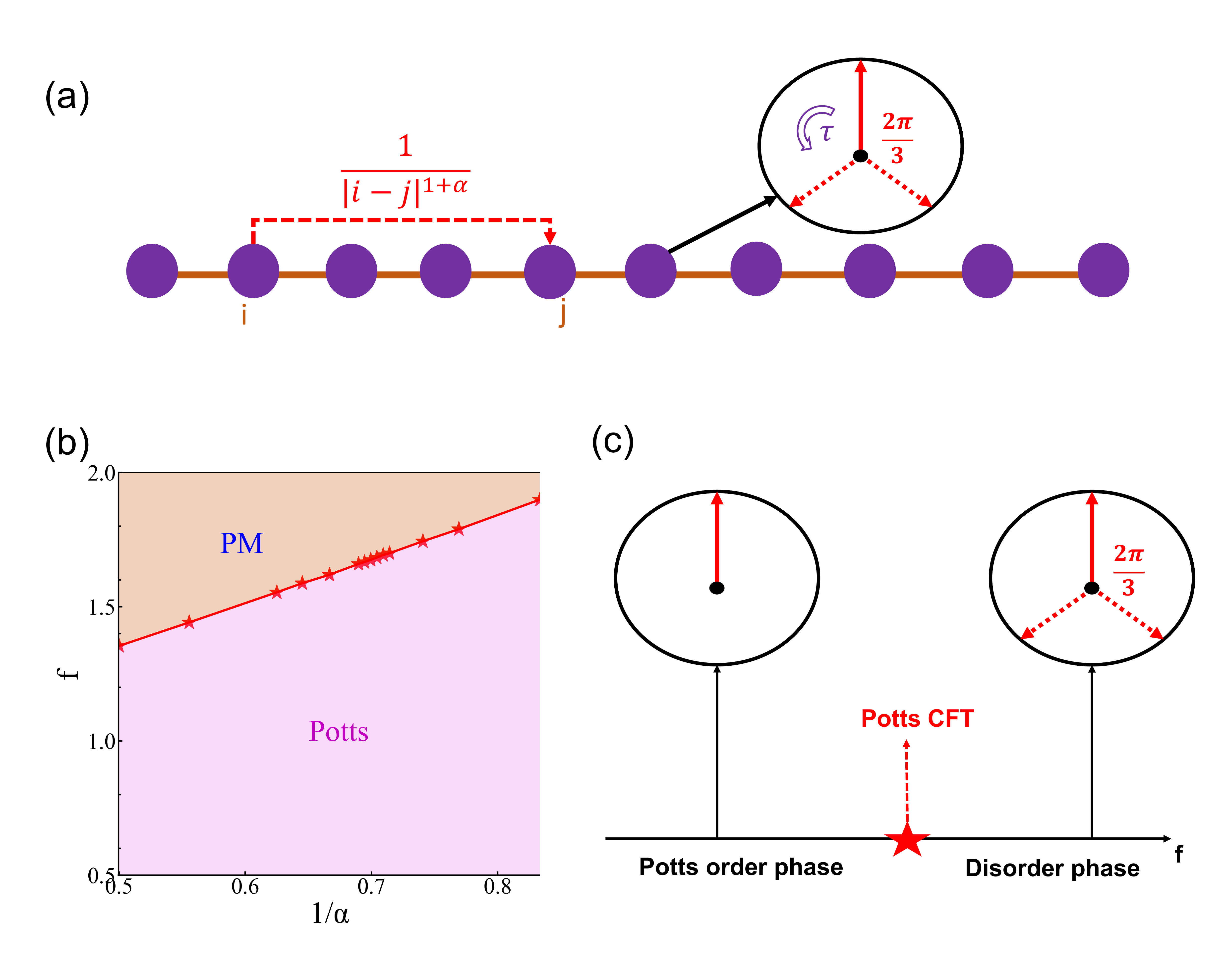}
\caption{(Color online) Schematic long-range interaction (a) and ground-state phase diagram with respect to $1/\alpha$ and external transverse field $f$ of quantum Potts chain with long-range interaction (b). In (b), Potts donates the Potts order phase. PM denotes the paramagnetic disorder phase (see the main text). The red line is the phase boundary between Potts and PM phase, and red star symbols denote the DMRG results of the critical values $f_{c}^{*}$. (c) The schematic phase diagram of standard quantum Potts chain with the nearest-neighboring interaction. The critical point between Potts ordered phase and disordered phase belongs to Potts universality class, which is described by Potts CFT.}
\label{fig:phase_diagram}
\end{figure}

The system of our study is a quantum three-state Potts chain with long-range power-law interactions (see Fig.~\ref{fig:phase_diagram}(a)), described by the following Hamiltonian~\cite{olyom1981prb,Huang2019prb}
\begin{equation}
\begin{split}
\label{E1}
&H_{LRP} = H_0 + fH_1 \\ 
&= -\frac{J}{N(\alpha)}\sum_{i,j}\frac{(\sigma^{\dagger}_{i}\sigma_{j}+\sigma_{i}\sigma^{\dagger}_{j})}{|i-j|^{d+\alpha}}-f\sum_{i}(\tau_{i}+\tau^{\dagger}_{i}),
\end{split}
\end{equation}
where $H_1$ and $H_0$ are the driving and undriving Hamiltonian, respectively. $J$ is the interaction strength, and $f$ represents the external transverse field, parameter $\alpha$ tunes the power of long-range interactions ($\frac{1}{|i-j|^{d+\alpha}}$), and $d$ is the spatial dimension (equal to 1 in our case). $N(\alpha)(= \frac{1}{N-1}\sum_{i,j,i\ne j}\frac{1}{r_{ij}^{\alpha}}$) is the Kac factor to preserve the Hamiltonian extensive. $\sigma$ dictates the direction of the watch hand, and $\tau$ rotates the watch hand clockwise through a discrete angle $2\pi/3$, as shown in Fig.~\ref{fig:phase_diagram}(a). $\sigma$ and $\tau$ satisfy $\sigma^{3}_{i}=I$,$\tau^{3}_{i}=I$, and $\sigma_{i}\tau_{j} = \omega \delta_{ij}\tau_{j}\sigma_{i}$, where $\omega = e^{2\pi i/3}$. A global $\mathbb{Z}_{3}$ transformation represented by $G = \prod_{i} \tau_{i} $ makes the Hamiltonian invariant. The operators are defined by 
\begin{equation}
\tau=
\begin{pmatrix}
1 & 0 & 0 \\
0 & \omega & 0 \\
0 & 0 & \omega^{2}
\end{pmatrix},\quad
\sigma=
\begin{pmatrix}
0 & 1 & 0 \\
0 & 0 & 1 \\
1 & 0 & 0
\end{pmatrix}.
\end{equation}

The system is in an ordered phase which breaks the $\mathbb{Z}_{3}$ symmetry for $f<<J$ and in a disordered paramagnetic phase (PM) for $f>>J$. The phase transition from the $\mathbb{Z}_{3}$-breaking Potts order to the $\mathbb{Z}_{3}$ symmetric disordered phase is described by the three-state Potts CFT with correlation length exponent $\nu=5/6$. The model becomes an infinite-range Potts chain (Lipkin–Meshkov–Glick model~\cite{LIPKIN1965188}) when $\alpha+d=0$, and a nearest neighbor quantum Potts chain when $\alpha+d =\infty$.

Non-perturbative numerical methods are employed to investigate critical behaviors of quantum Potts chains with long-range power-law interaction and estimate the value of $\alpha_c$ in the "non-classical" region. Inspired by previous RG results 
for classical systems with long-range interaction~\cite{defenu2109}, only parameter region $1.2 \le \alpha \le 2.0$ is considered. Considering that the quantum Potts chain with long-range interaction do not have exact solutions in the parameter region of interest, a large-scale finite-size DMRG method~\cite{White1992prl,Schollw2005rmp,SCHOLLWOCK201196} based on MPS~\cite{SCHOLLWOCK201196,Verstraete2004prl}, which is one of the most powerful numerical method for one-dimensional strongly correlated many-body systems, is employed. The MPS bond dimension is set to $300$; good convergence of true energy eigenstates and fidelity susceptibilities are guaranteed by requiring relative energy errors less than $10^{-8}$. The fidelity susceptibility defined in Eq.~\ref{E2} is computed with a minimal step $\delta f = 10^{-3}$. The strength of the interaction $J=1$ as an energy unit, and open boundary conditions are applied. 

\subsection{Fidelity susceptibility and scaling relations}%
\label{sec:FS}
The system undergoes a continuous phase transition from an ordered to a disordered phase when tuning the external field $f$ to a critical value $f_{c}^{*}$, at which the structure of the ground state wave function change significantly. The quantum ground-state fidelity $F(f,f + \delta f)$, defined as the overlapping amplitude of the ground state wave function with the external field $f$ and the ground state wave function with the external field $f+\delta f$~\cite{gu2010fidelity,Gu2009pre,Gu_2009,You_2015,Gu_2014,damski2016fidelity}, and its value is almost zero near $f_{c}^{*}$, that is, $F(f_{c}^{*},f_{c}^{*} + \delta f)\sim 0$.
 In practice, the more convenient quantity to characterize QPTs is the fidelity susceptibility, defined by the leading term of the fidelity,
\begin{equation}
\begin{split}
\label{E2}
\chi_{F}(f)={\rm{lim}}_{\delta f \rightarrow 0}\frac{2(1-F(f,f+\delta f))}{(\delta f)^{2}}.
\end{split}
\end{equation} 

For a continuous quantum phase transition of a finite system with size $L$, fidelity susceptibility exhibits a peak at pseudo-critical point $f_c(L)$, and the value of the quantum critical point $f_{c}^{*}$ can be estimated by polynomial fitting $f_{c}(L)=f_{c}^{*}+aL^{-b}$~\cite{sandvik2010computational}. In the vicinity of $f_{c}^{*}$, previous studies~\cite{Gu_2014,gu2010fidelity,Yu2014,Albuquerque2010prb,Schwandt2009prl,Yu2009pre,sun2019prb,sun2015prb,onig2016prb,Albuquerque2010prb,gu2010fidelity} have shown that the finite-size scaling behaviors of fidelity susceptibility $\chi_{F}(f)$ follows 
\begin{equation}
\begin{split}
\label{E4}
\chi_{F}(f\rightarrow f_{c}^{*})\propto L^{\mu} 
\end{split}
\end{equation}
and
\begin{equation}
\begin{split}
\label{E5}
L^{-d}\chi_{F}(f) = L^{(2\nu)-d}f_{\chi_{F}}(L^{1/\nu}|f-f_{c}^{*}|),
\end{split}
\end{equation}
where $\mu (=2+2z-2 \Delta_{V} )$ is the critical adiabatic dimension~\cite{gu2010fidelity}, $z$ is dynamic exponent, $\Delta_{V}$ is the scaling dimension of the local interaction $V(x)$ at $f_{c}^{*}$,  $\nu$ is the critical exponent of the correlation length, $d$ is the spatial dimension of the system, and $f_{\chi_{F}}$ is an unknown scaling function. 
Based on Eq.~\ref{E4} and.~\ref{E5}, the values of critical exponents $\nu$ and $\mu$ of the QPT can be determined, and the universality class to which the QPT belongs can also be determined. Note that in practice, the critical exponent $\mu$ is usually extracted from fidelity susceptibility per site, $\chi_{L}(f) = \chi_{F}(f)/L^{d}$.

\section{PHASE DIAGRAM AND CRITICAL BEHAVIOR}
\label{sec:phase}

\subsection{Quantum phase diagram}

The ground state phase diagram of the quantum Potts chain with long-range interactions for $\alpha > 0$ (Eq.~\ref{E1}) is obtained by performing large-scale DMRG simulations with $L=96,120,144,156,168,192,216,240$ sites.
The result is presented in Fig.~\ref{fig:phase_diagram}(b). For $\alpha \rightarrow \infty $, the ground state is a Potts order phase with three-fold degeneracy for $f=0$ and a paramagnetic disorder phase for $ f \rightarrow \infty$, consistent with previous results~\cite{Yu_cbc_2021}(also see Fig.~\ref{fig:phase_diagram}(c)). Furthermore, for finite $\alpha$, it is found that the quantum Potts chain with long-range interactions has a stable Potts order and a disordered phase over the entire range of $\alpha$ we investigate.

\begin{figure}[tb]
\includegraphics[width=0.5\textwidth]{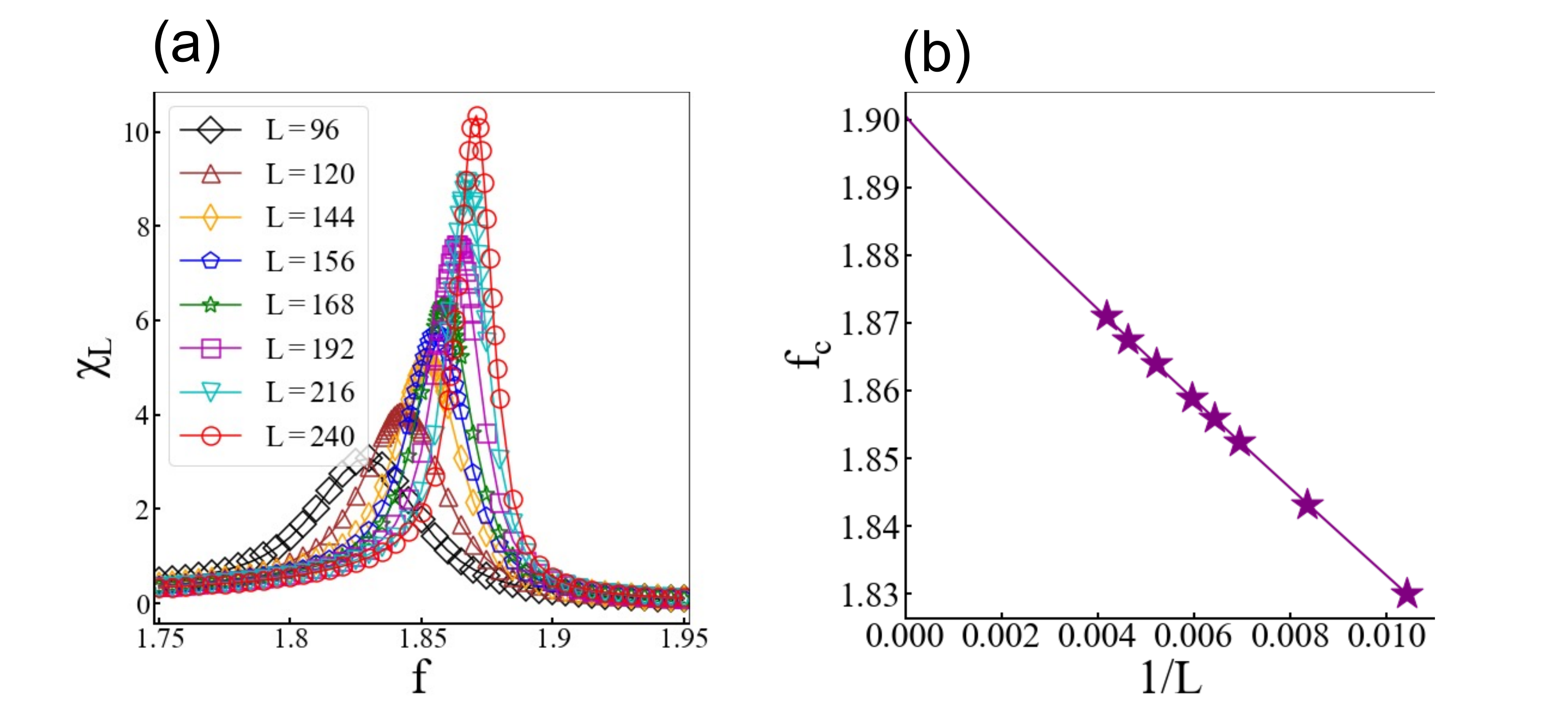}
\caption{(Color online) (a) Fidelity susceptibility per site $\chi_{L}$ of the Potts chain with long-range interaction for $\alpha=1.2$ and $L=96,120,144,156,168,192,216,240$ sites as a function of external transverse field $f$; symbols denote finite-size DMRG results. (b) Extrapolation of critical point $f_{c}^{*}$ for the Potts chain with long-range interaction; symbols denote the finite-size DMRG results for $\alpha=1.2$ and $L=96,120,144,156,168,192,216,240$ sites. We use polynomial fitting $f_{c}(L) = f_{c}^{*} + aL^{-b}$ and extrapolate the critical point $f_{c}^{*}$ = 1.89878.}
\label{fig:fs_fc}
\end{figure}

The finite-size scaling behavior of fidelity susceptibility for $\alpha = 1.2$ with different $L$
 is presented in Figure~\ref{fig:fs_fc}(a), which 
obeys $\chi_{L}(f_{c}^{*})\propto L^{\mu-1}$ (Eq.~\ref{E4}) near the second-order QPT critical point. As system size $L$ increases,
the peak position $f_{c}(L)$ gets closer and closer to the exact critical point value $f_{c}^{*}$. More precisely, 
for the long-range interaction Potts chain with $\alpha=1.2$, $f_{c}^{*}$ is determined by polynomial fitting $f_{c}(L) = f_{c}^{*} + aL^{-b} $, and then extrapolating to $L$ to infinity (Fig.~\ref{fig:fs_fc}(b)). According to Eq.~\ref{E5}, the fidelity susceptibility follows an exact scaling relation, and collapses to one master curve (Fig.~\ref{fig:mu_data}(b)), confirming that the extrapolation is appropriate. The finite-size scaling behavior of fidelity susceptibility for other $\alpha$ is also investigated (see Appendix~\ref{sec:A1}), and the results are presented in Table~\ref{tab:exponents}. Results show that the quantum critical point moves to lower $f_{c}^{*}$ values as $\alpha$ increases.

\subsection{Finite-size scaling and critical exponent}

\begin{figure}[tb]
\includegraphics[width=0.5\textwidth]{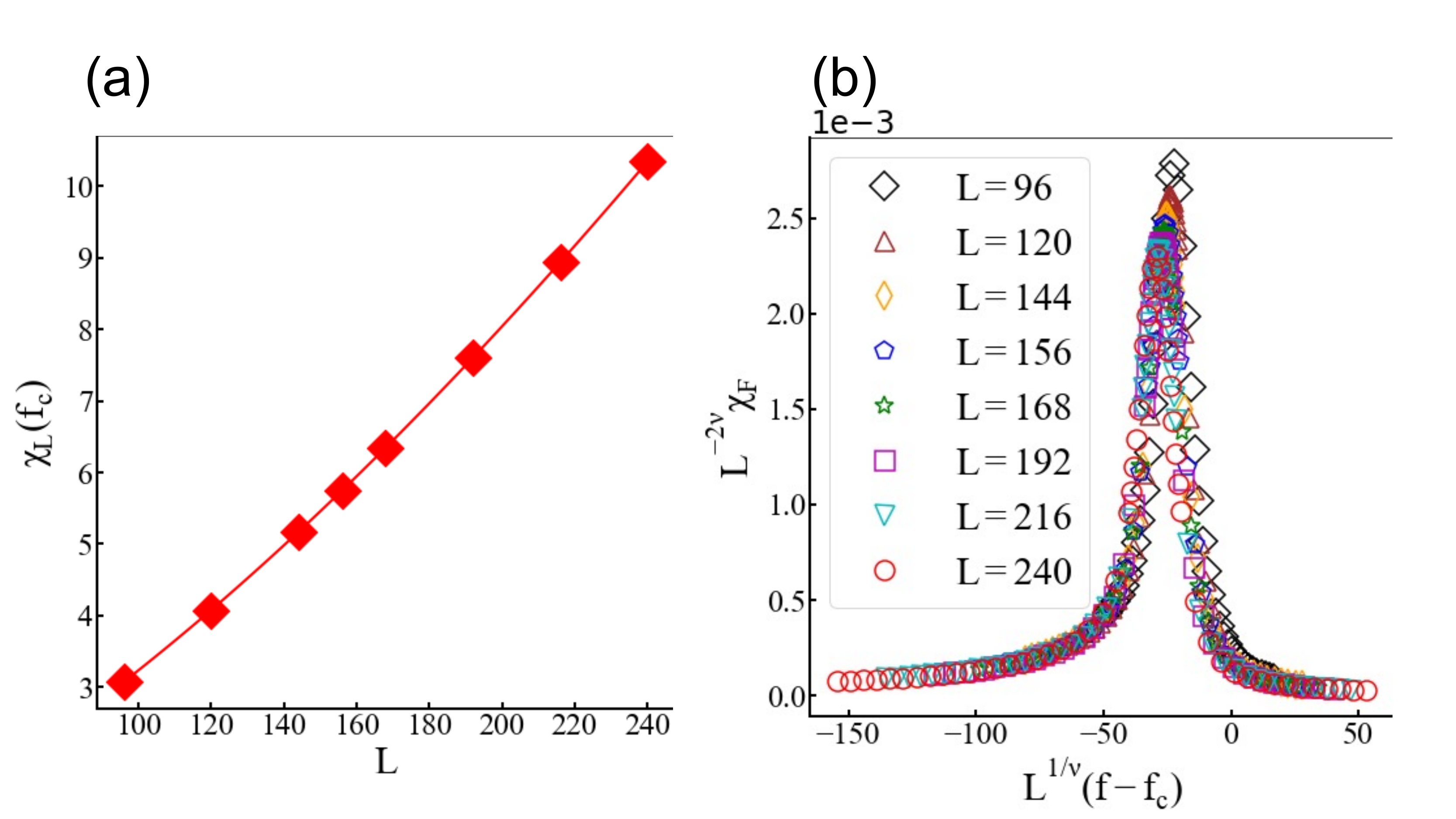}
\caption{(Color online) (a) The maximal of fidelity susceptibility per site $\chi_{L}(f_{c}^{*})=\chi_{F}(f_{c}^{*})/L$ as a function of system sizes $L$ for $\alpha=1.2$. We use polynomial fitting $\chi_{L} \sim L^{\mu}(a+bL^{-1})$ and extrapolate the critical adiabatic dimension $\mu$ = 2.53501. (b) Data collapse of fidelity susceptibility $\chi_{F}$ for the Potts chain with long-range interaction; symbols denote the finite-size DMRG results for $\alpha=1.2$ and $L=96,120,144,156,168,192,216,240$ sites, where $\nu=0.78895$ and $f_{c}^{*}=1.89878$ are used for data collapse plots.}
\label{fig:mu_data}
\end{figure}

The next questions are what is the critical behavior of the long-range interaction Potts chains with different $\alpha$ values, and whether there is a critical threshold $\alpha_c$, at which the critical behavior changes continuously from a long-range universality class to a short-range one? To this end, we calculate the critical exponents $\mu$ and $\nu$ of the fidelity susceptibility in the region $1.2 \le \alpha \le 2.0$ based on large-scale DMRG simulations for different $L$. The value of the fidelity susceptibility per site, $\chi_{L}=\chi_{F}/L$, at the peak position $f_{c}(L)=1.89878$ for different $L$ at $\alpha=1.2$ is shown in Fig.~\ref{fig:mu_data}(a). It can be well fitted by a polynomial fitting of $\chi_{L} \sim L^{\mu}(a+bL^{-1})$. According to Eq.~\ref{E4}, the adiabatic critical dimension $\mu = 2.53501$ is then obtained by extrapolation $L$ to infinity.
 
 

 
 
\begin{figure}[tb]
\includegraphics[width=0.48\textwidth]{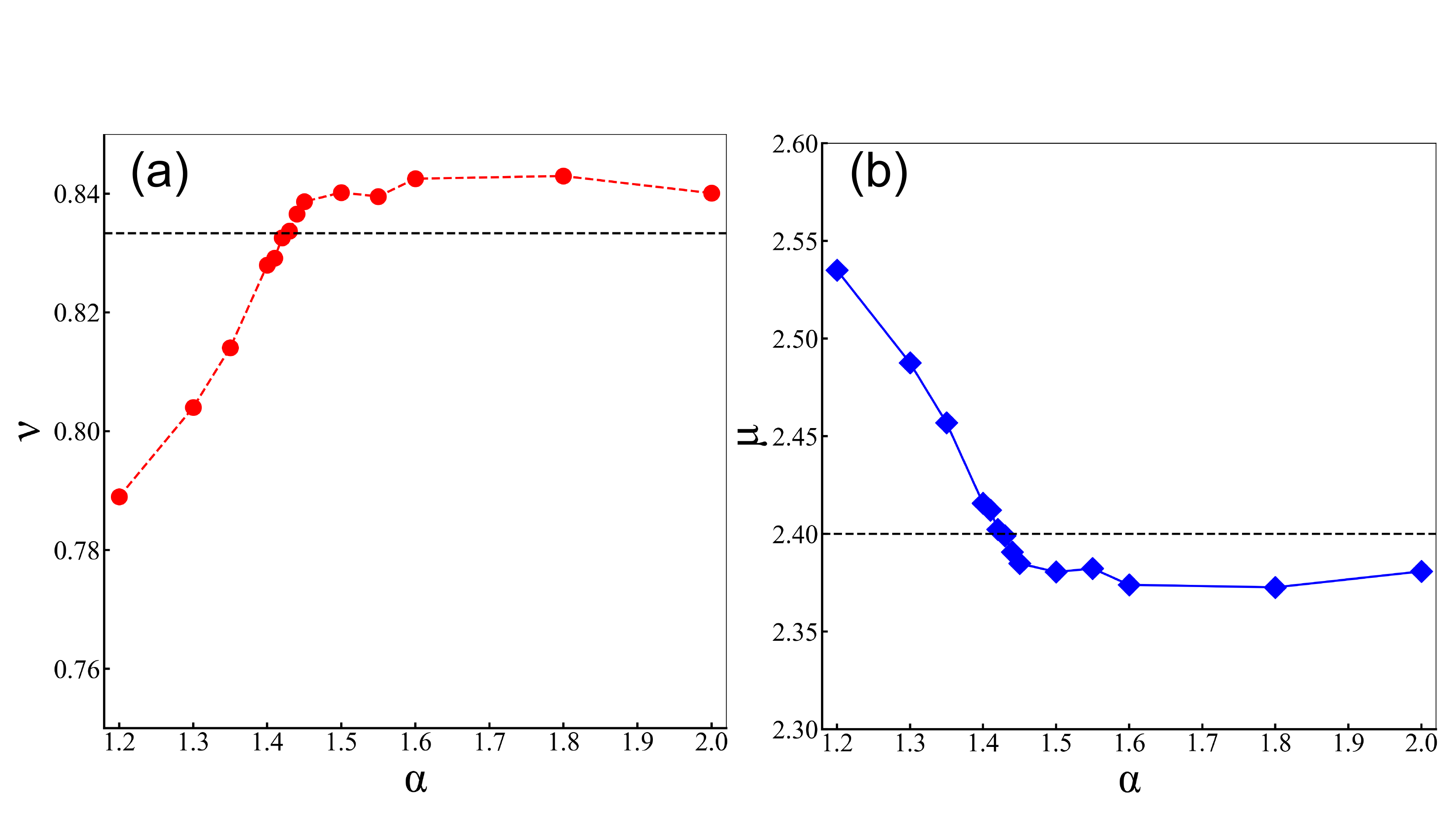}
\caption{(Color online) Critical exponent of the correlation length $\nu$ (black dash line refers to 2D three-state Potts correlation length exponent $\nu=5/6$ as a comparison) (a) and  critical adiabatic dimension $\mu$ (black dash line refers to 2D three-state Potts critical adiabatic dimension $\mu=12/5$ as a comparison) (b) with respect to $\alpha$ for the Potts chain with long-range interaction; the symbols denote the finite-size DMRG results that are obtained by extrapolating from the fidelity susceptibility $\chi_{F}(f_{c}^{*})$ at the peak position $f_{c}^{*}$ of $L = 96, 120,144,156,168,192,216,240$ sites.}
\label{fig:fig6}
\end{figure}

 According to Eq.~\ref{E5}, the fidelity susceptibility can be scaled by $L^{-2/\nu}\chi_{F}$ as a function of $L^{\nu}(f-f_{c}^{*})$ in the vicinity of the quantum critical point $f_{c}^{*}$. The critical correlation length exponent $\nu$(=0.78895) is then determined as the value at which all fidelity susceptibilities for different $L$ collapse into a single one (Fig.~\ref{fig:mu_data}(b)). The calculations of the critical adiabatic dimension $\mu$ and the correlation length exponent $\nu$ for other $\alpha$ are presented in Appendix~\ref{sec:A3}, and the results of all $\alpha$ are summarized in Table.~\ref{tab:exponents}. As can be seen from  Fig.~\ref{fig:fig6}(a), either $\nu$ or  $\mu$ as function of $\alpha$ shows a crossover at $\alpha=\alpha_c=1.43$. When $\alpha<\alpha_c$, $\mu$ and $\nu$ are monotonic functions of $\alpha$. In contrast, when $\alpha>\alpha_c$, they are more or less constant and approach the critical exponent values of the 2D three-state Potts model, $\nu=5/6$ and $\mu=12/5$, respectively, within $0.8\%$ error due to finite size effect  (black dash line in Fig.~\ref{fig:fig6}). Therefore, the critical behavior of the fidelity susceptibility undergoes a continuous crossover at $\alpha_c\approx 1.43$, from the long-range universality class region with varying correlation length exponent ($\alpha\textless \alpha_c$) to the short-range universality class region with constant exponents ($\alpha\textless \alpha_c$, three-state Potts region). This tendency is different from $\mathbb{Z}_{2}$ symmetric (Ising) quantum spin chain with long-range interaction~\cite{sun2017pra,sun2018pra}.


\begin{ruledtabular}
\begin{table}[tb]
\caption{Critical exponents of the Potts chain with long-range interaction for different $\alpha$. Critical exponents in the standard quantum Potts chain($\alpha=\infty$) are also listed for comparison. The critical threshold of long-range interaction power $\alpha_c \sim 1.43$.}
\label{tab:exponents}
\begin{tabular}{cccc}
$\alpha$	& $f_{c}^{*}$			& $\nu$	& $\mu$	\\
\hline
1.2	& 1.89878				& 0.78895			& 2.53501			\\
1.3		& 1.78880				& 0.80400			& 2.48750		\\
1.35		& 1.74389					& 0.81406			& 2.45681						\\
1.4		& 1.69934					& 0.82796					& 2.41558						\\
1.41         & 1.69227                                  & 0.82912                                  & 2.41220            \\
1.42         & 1.68301                                  & 0.83255                                  & 2.40224            \\
1.43          &1.67415                                  & 0.83367                                  & 2.39903            \\
1.44          &1.66635                                  & 0.83658                                  & 2.39069           \\
1.45          &1.65914                                  & 0.83864                                  & 2.38481           \\
1.5		& 1.61911					& 0.84018					& 2.38045						\\
1.55		& 1.58752					& 0.83951					& 2.38223						\\
1.6		& 1.55358				& 0.84251					& 2.37386						\\
1.8		& 1.44159				& 0.84296				& 2.37258						\\
2.0		& 1.35406				& 0.84007				& 2.38076						\\
$\infty$		&1.00000				& 0.83333				& 2.40000						\\
\end{tabular}
\end{table}
\end{ruledtabular}

\subsection{Discussion}


The application of RG techniques to {\it{classical}} spin systems with long-range interactions provides a good understanding of phase transitions that occur within them. Perturbative two-loop RG calculations show that the three-state Potts chain with long-range power-law interactions has three parameter regimes, (1) small $\alpha$ ($\alpha \textless d/2$), (2) intermediate $\alpha$ ($ d/2 \textless \alpha \textless \alpha_c $, and (3) large $\alpha$ ($\alpha \textgreater \alpha_c$) regions, similar to the classical three-state Potts model at low-energy and long-distance that can be described by $\phi^{3}$ Landau-Ginzberg-Wilson effective action~\cite{defenu2109,Theumann1985prb}. Moreover, previous theoretical and numerical results~\cite{Defenu2017prb,sun2017pra,sun2018pra,Fisher1972prl,Defenu2017prb} show that the critical behavior of ${\rm{O(N)}}$ symmetric {\it {quantum}} model systems with long-range interactions is consistent with that of {\it {classical}} ${\rm{O(N)}}$ ones with effective dimension $d_{{\rm{eff}}} (= \frac{2d}{\alpha} + 1)$ for $d/2 \textless \alpha \textless \alpha_c$. However, for {\it quantum} systems with long-range interactions in the {\it "non-classical" region} ($d/2 \textless \alpha \textless \alpha_c $), the quantum-classical correspondence is very subtle and there is no analytical expression for the critical exponents. Particularly, it is unclear whether the critical behavior of the $\mathbb{Z}_{3}$ symmetric {\it quantum} spin systems with long-range interactions is also consistent with the two-loop RG results of the {\it classical} Potts model with long-range interactions.

For {\it quantum} three-state Potts model systems with long-range power-law interactions, using non-perturbative DMRG, we found that there exists a critical value $\alpha_c$ in the long-range power-law interactions. When $\alpha \textless \alpha_{c}$, the RG flow ends in a stable long-range fixed point with varying critical exponents $\nu$, and when $ \alpha \textgreater \alpha_{c}$, it ends in a short-range fixed point. These results for {\it quantum} three-state Potts model systems with long-range power-law interactions are in qualitative agreement with previous two-loop RG results for {\it classical} $\phi^{3}$ theory~\cite{Theumann1985prb}. More importantly, for the first time, we numerically determine the critical threshold of the long-range interaction power $\alpha_{c} \approx 1.43$ in a non-perturbative manner, which is more accurate than previous perturbative two-loop RG results $\alpha_c \sim 1.73$.


\section{CONCLUSION}%
\label{sec:summary}
To summarize, we investigate the critical behavior of quantum Potts chain with long-range interactions through large-scale DMRG simulations. Using the fidelity susceptibility as a diagnostic, we obtain a ground-state phase diagram between PM and Potts order phases. As the long-range interaction power increases, the location of the quantum critical point shifts to weaker external fields. The finite-size scaling of the fidelity susceptibility $\chi_{F}$ and the nature of the QPTs of the quantum Potts chain with long-range interaction are also investigated. Our numerical results show that there is a critical threshold in the long-range interaction power, long-range fixed points are stable for $\alpha \textless \alpha_{c}$ and short-range fixed points are stable for $ \alpha \textgreater \alpha_{c}$. These results are consistent with previous two-loop RG calculations from classical $\phi^{3}$ theory, but differ from the quantum Ising chains with long-range interaction. In addition, for the first time, we determined the critical long-range interaction power $\alpha_c \approx 1.43$ in a non-perturbative way, which is more precise than the previous perturbative two-loop RG results $\alpha_c \sim 1.73$. Interesting future questions include the fate of finite temperature effects in quantum Potts chains with long-range interaction, and the critical behavior of quantum four-state Potts models with long-range power interactions. Our work could shed new light on the interplay between long-range interactions (frustrated) and many-body physics.

\begin{acknowledgements}
We thank Sheng Yang, and Nicolo Defenu for helpful discussions and communication. Numerical simulations were carried out with the ITensor package~\cite{itensor2007}. We also thank the computational resources provided by the TianHe-1A supercomputer, the High Performance Computing Platform of Peking University, China. This work is supported by National Natural Science Foundation of China under Grant No.11935002, and the National 973 project under Grant No. 2021YF1400501. C.D. was supported by the National Science
Foundation of China under Grants No. 11975024, the Anhui Provincial Supporting Program for Excellent Young Talents in Colleges and Universities under Grant No. gxyqZD2019023.
\end{acknowledgements}
  
\bibliography{LRPotts_qpt_arxiv}

\begin{appendix}

\section{FIDELITY SUSCEPTIBILITY FOR OTHER INTERACTION POWERS}
\label{sec:A1}

In this section, we provide additional data to show fidelity susceptibility for other interaction powers. 

\begin{figure*}[tb]
\includegraphics[width=0.96\textwidth]{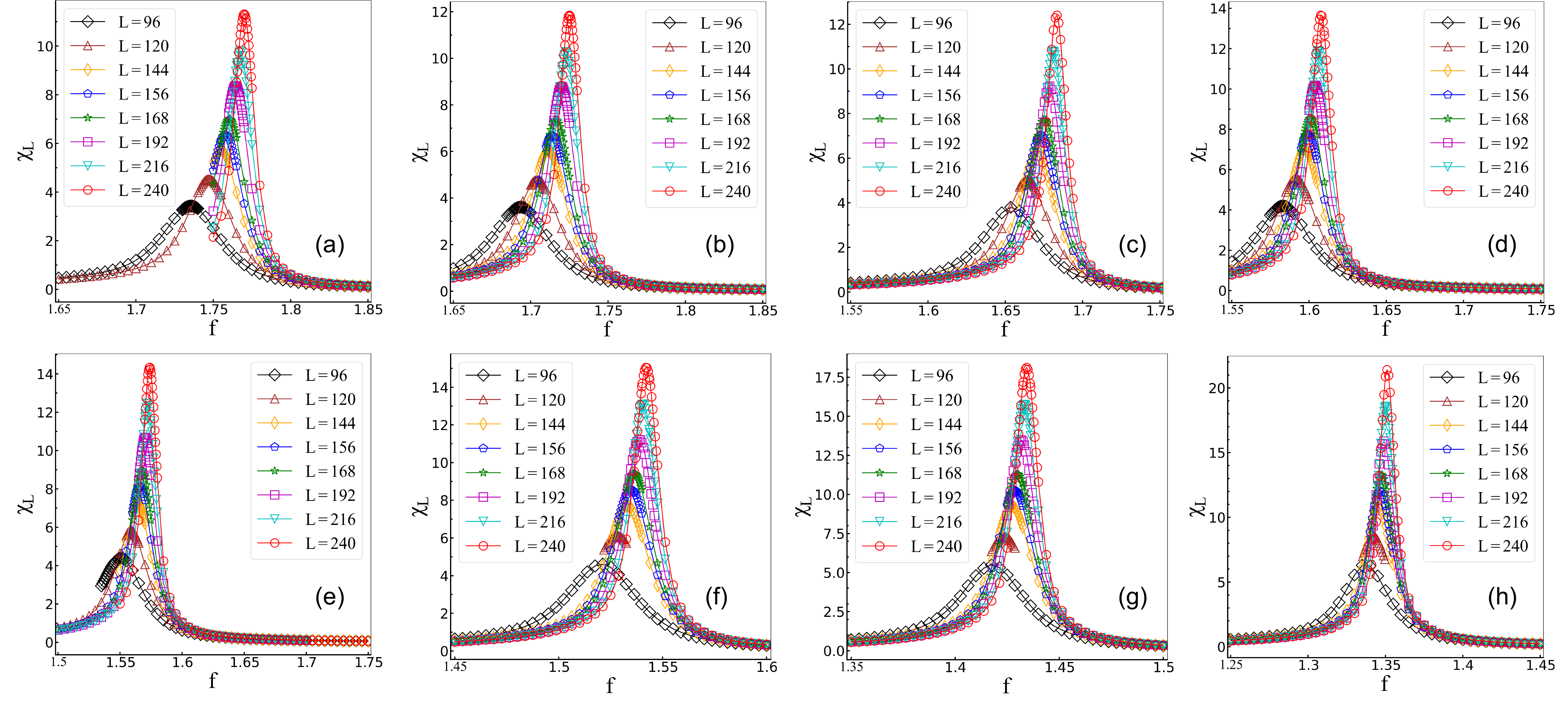}
\caption{(Color online) Fidelity susceptibility per site $\chi_{L}$ of the Potts chain with long-range interaction for (a) $\alpha=1.3$,(b) $\alpha=1.35$,(c) $\alpha=1.4$, (d) $\alpha=1.5$, (e) $\alpha=1.55$, (f) $\alpha = 1.6$, (g) $\alpha=1.8$, (h) $\alpha = 2.0$, and $L=96,120,144,156,168,192,216,240$ sites as a function of external transverse field $f$; symbols denote finite-size DMRG results.}
\label{fig:appA}
\end{figure*}

\begin{figure*}[tb]
\includegraphics[width=0.96\textwidth]{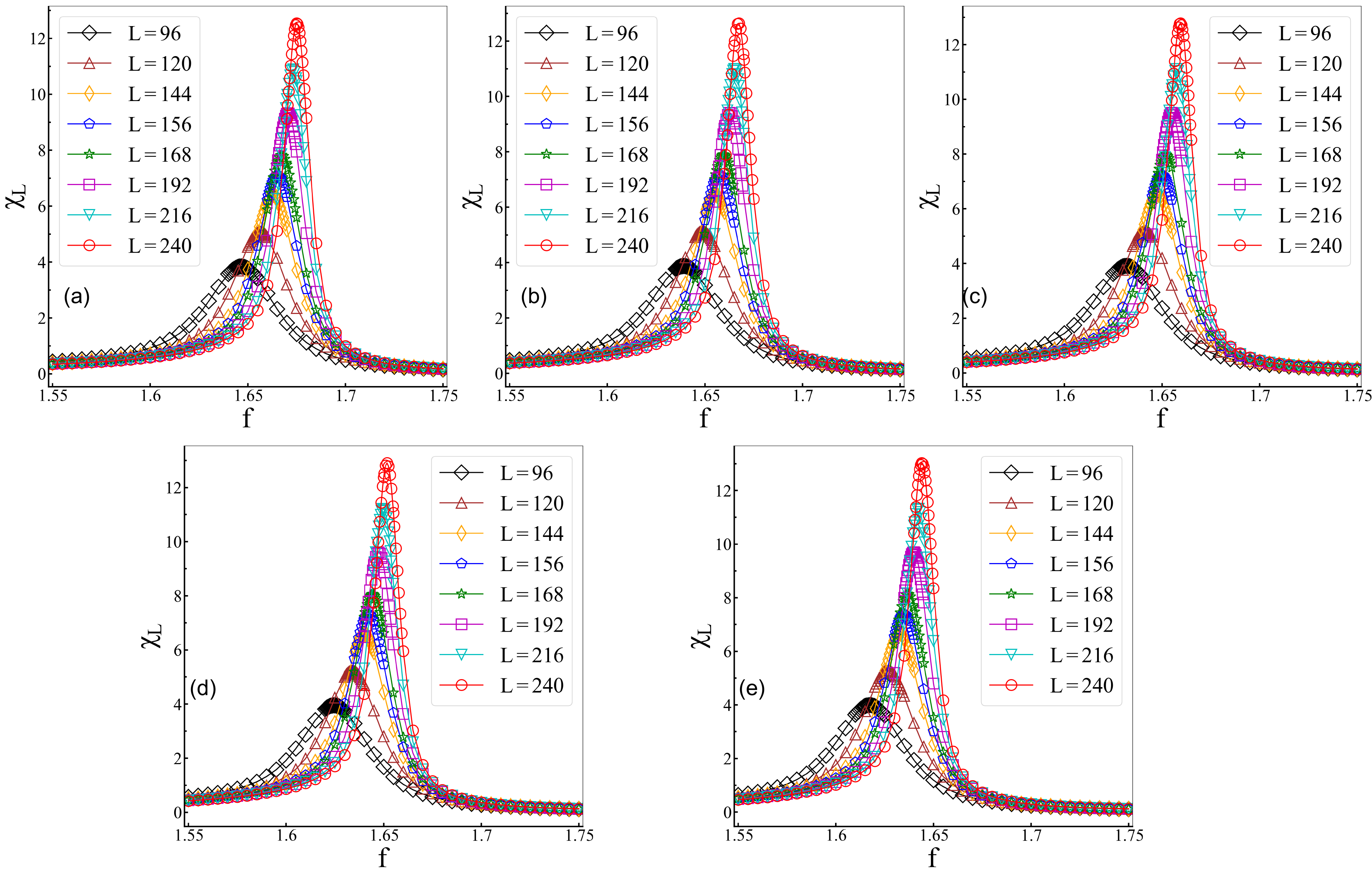}
\caption{(Color online) Fidelity susceptibility per site $\chi_{L}$ of the Potts chain with long-range interaction for (a) $\alpha=1.41$, (b) $\alpha=1.42$, (c) $\alpha=1.43$, (d) $\alpha=1.44$, (e) $\alpha=1.45$, and $L=96,120,144,156,168,192,216,240$ sites as a function of external transverse field $f$; symbols denote finite-size DMRG results.}
\label{fig:appA2}
\end{figure*}

As the same in the main text, on the one hand, fidelity susceptibility per site $\chi_{L}$ of the Potts chain with long-range interaction for $\alpha=1.3$ (a), $\alpha=1.35$ (b), $\alpha=1.4$ (c), $\alpha=1.5$ (d), $\alpha=1.55$ (e), $\alpha=1.6$ (f), $\alpha=1.8$ (g), $\alpha = 2.0$ (h), and $L=96,120,144,156,168,192,216,240$ sites as a function of external transverse field $f$, are shown in the Fig.~\ref{fig:appA}. On the other hand, in order to determine critical $\alpha_c$, we also show fidelity susceptibility per site for $\alpha = 1.41$ (a), $\alpha=1.42$ (b), $\alpha=1.43$ (c), $\alpha=1.44$ (d), $\alpha=1.45$ (e) in the Fig.~\ref{fig:appA2}. We find that quantum critical points are shifted to lower values of $f$ as long-range interaction power increases.

\section{DATA COLLAPES FOR OTHER INTERACTION POWERS}
\label{sec:A2}

\begin{figure*}[tb]
\includegraphics[width=0.96\textwidth]{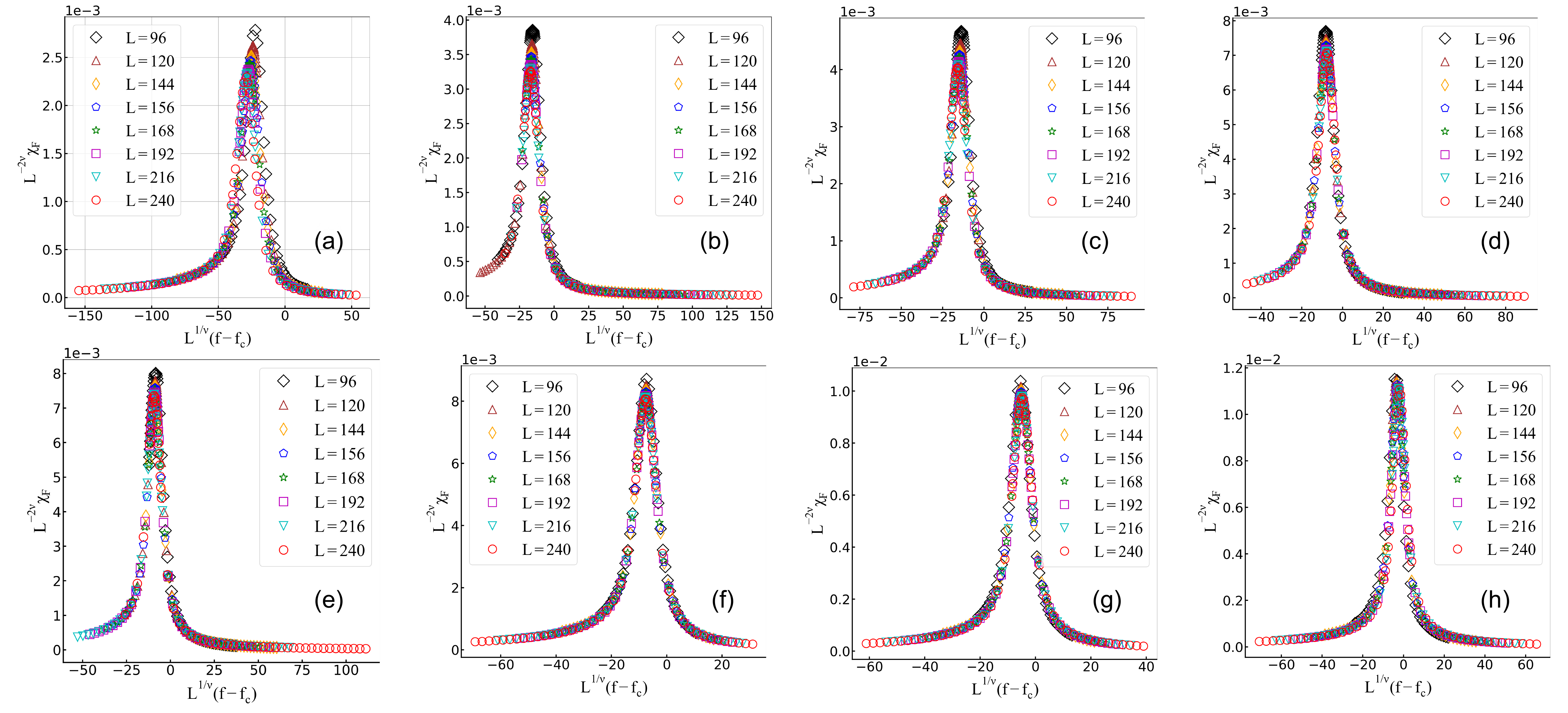}
\caption{(Color online) Data collapse of fidelity susceptibility $\chi_{F}$ for the Potts chain with long-range interaction; symbols denote the finite-size DMRG results for (a) $\alpha=1.3$, (b) $\alpha=1.35$,  (c) $\alpha=1.4$, (d) $\alpha=1.5$,  (e) $\alpha=1.55$, (f) $\alpha=1.6$, (g) $\alpha=1.8$, (h) $\alpha=2.0$, where the varying tendency of correlation length exponents in the "non-classical" region is consistent with the theoretical analysis.}
\label{fig:appB}
\end{figure*}

In this section, we provide additional data to show that the varying tendency of long-range correlation length exponents in the "non-classical" region is consistent with theoretical analysis. 

\begin{figure*}[tb]
\includegraphics[width=0.96\textwidth]{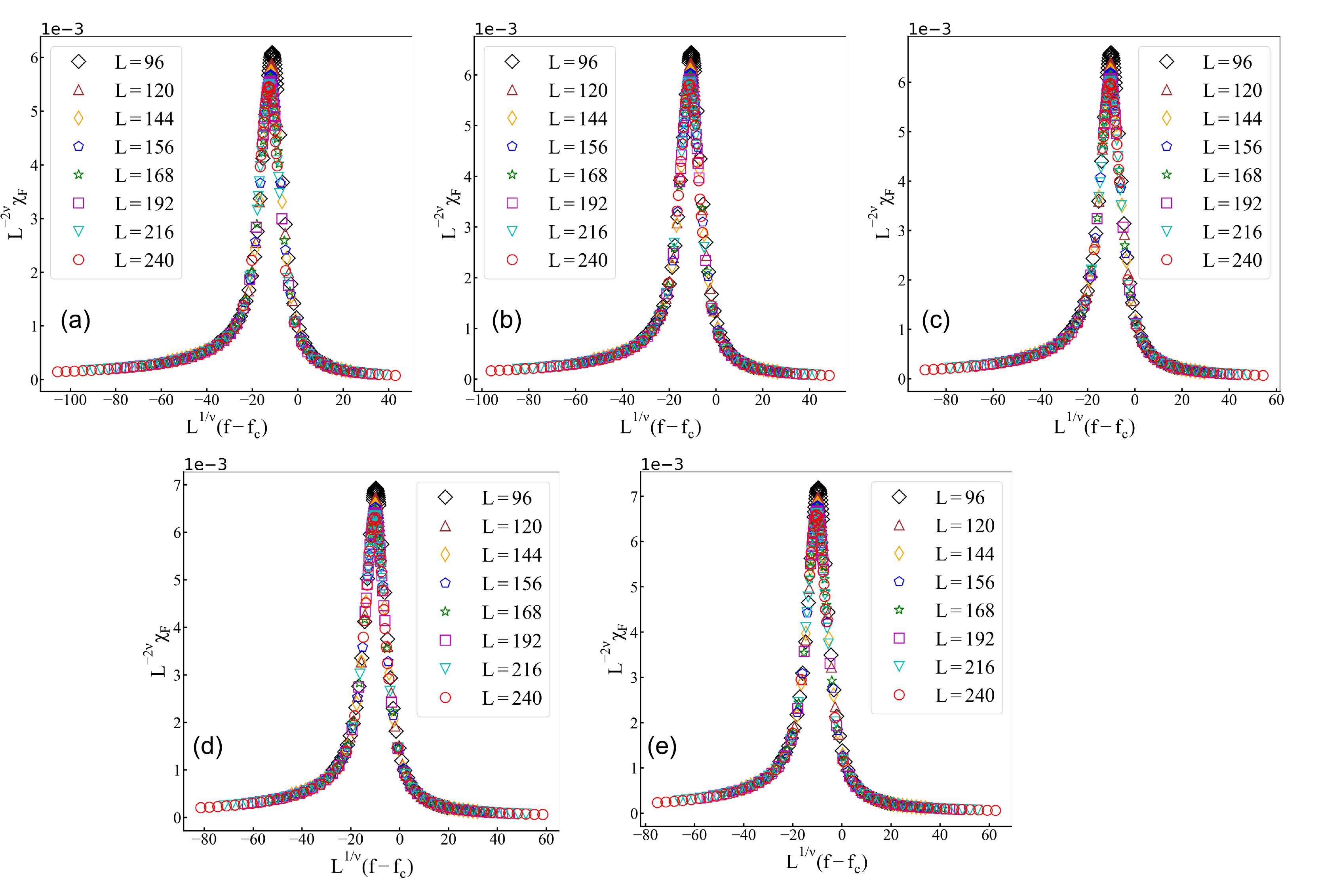}
\caption{(Color online) Data collapse of fidelity susceptibility $\chi_{F}$ for the Potts chain with long-range interaction; symbols denote the finite-size DMRG results for (a) $\alpha=1.41$, (b) $\alpha=1.42$, (c) $\alpha=1.43$, (d) $\alpha=1.44$, (e) $\alpha=1.45$, where the varying tendency of correlation length exponents in the "non-classical" region is consistent with the theoretical analysis.}
\label{fig:appB2}
\end{figure*}

As the same in the main text, on the one hand, data collapse of fidelity susceptibility $chi_{F}$ for the Potts chain with long-range interaction, $\alpha=1.3$ (a), $\alpha=1.35$(b), $\alpha=1.4$ (c),  $\alpha=1.5$ (d), $\alpha=1.55$ (e), $\alpha=1.6$ (f), $\alpha=1.8$ (g), $\alpha=2.0$(h), and $L=96,120,144,156,168,192,215,240$ sites, are shown in Fig.~\ref{fig:appB}. On the other hand, in order to determine critical $\alpha_c$, we also show data collapse for $\alpha = 1.41$ (a), $\alpha=1.42$ (b), $\alpha=1.43$ (c), $\alpha=1.44$ (d),$\alpha=1.45$ (e) in the Fig.~\ref{fig:appB2}. The correlation length exponents are summarized in Table.~\ref{tab:exponents}. We clearly see that the varying tendency of correlation length exponents in the "non-classical" region is consistent with the theoretical analysis.

\section{QUANTUM ADIABATIC DIMENSION FITTING FOR OTHER INTERACTION POWERS}
\label{sec:A3}

\begin{figure*}[tb]
\includegraphics[width=0.96\textwidth]{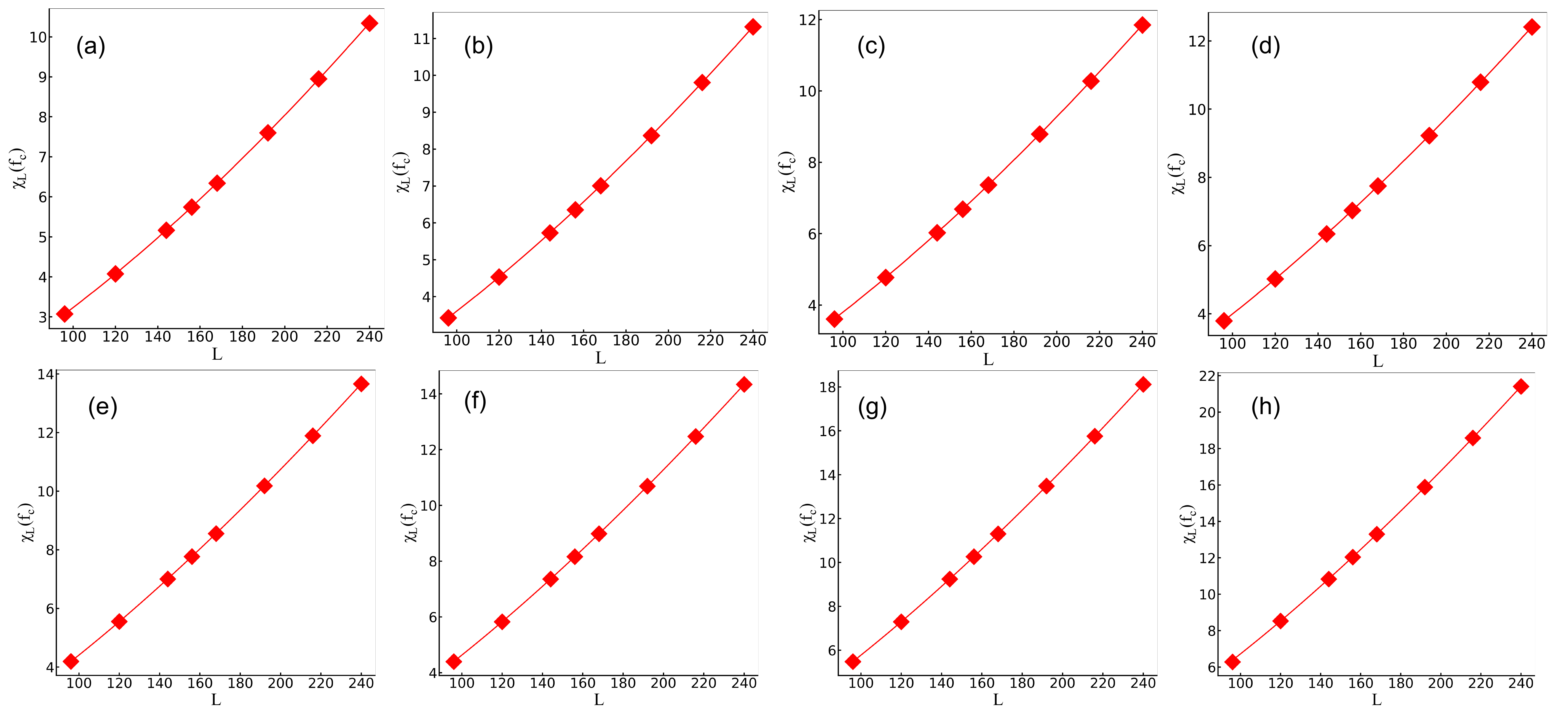}
\caption{(Color online) The maximal of fidelity susceptibility per site $\chi_{L}(f_{c}^{*})=\chi_{F}(f_{c}^{*})/L$ as a function of system sizes $L$ for (a) $\alpha=1.3$, (b) $\alpha=1.35$, (c) $\alpha=1.4$, (d) $\alpha=1.5$, (e) $\alpha=1.55$, (f) $\alpha=1.6$, (g) $\alpha=1.8$, (h) $\alpha=2.0$. We use polynomial fitting formula: $\chi_{L}(f_{c}^{*})=L^{\mu}(a+bL^{-1})$}
\label{fig:appC}
\end{figure*}

In this section, we provide additional data to extrapolate critical adiabatic dimensions for other long-range interaction powers. 

\begin{figure*}[tb]
\includegraphics[width=0.96\textwidth]{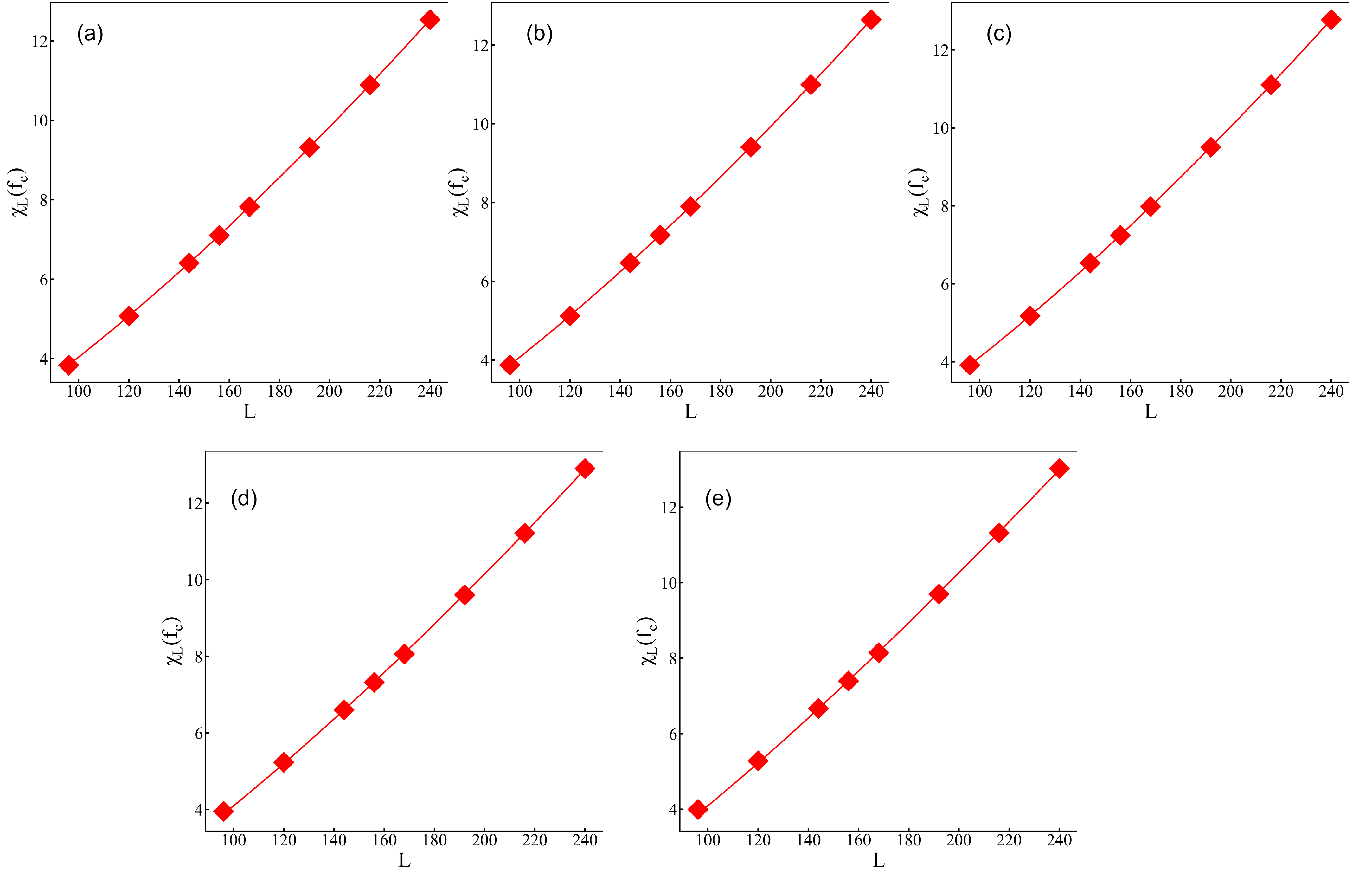}
\caption{(Color online) The maximal of fidelity susceptibility per site as a function of system sizes $L$ for (a) $\alpha=1.41$,(b) $\alpha=1.42$,(c) $\alpha=1.43$, (d) $\alpha=1.43$, (e) $\alpha=1.45$. We use polynomial fitting formula: $\chi_{L}(f_{c}^{*})=L^{\mu}(a+bL^{-1})$}
\label{fig:appC2}
\end{figure*}

As the same in the main text, on the one hand, the maximal of fidelity susceptibility per site $\chi_{L}(f_{c}^{*})=\chi_{F}(f_{c}^{*})/L$ as a function of system sizes $L$ for $\alpha=1.3$ (a), $\alpha=1.35$ (b), $\alpha=1.4$ (c),  $\alpha=1.5$ (d), $\alpha=1.55$ (e), $\alpha=1.6$ (f), $\alpha=1.8$ (g), $\alpha=2.0$ (h), and $L=96,120,144,156,168,192,216, 240$ sites, are shown in Fig.~\ref{fig:appC}. On the other hand, in order to determine critical $\alpha_c$, we also show the maximal of fidelity susceptibility per site for $\alpha = 1.41$ (a), $\alpha=1.42$ (b), $\alpha=1.43$ (c), $\alpha=1.44$ (d),$\alpha=1.45$ (e) in the Fig.~\ref{fig:appC2}. The critical adiabatic dimensions are summarized in the Table.~\ref{tab:exponents}. We clearly see that the varying tendency of critical adiabatic dimension in the "non-classical" region is consistent with the theoretical analysis.

\section{QUANTUM CRITICAL POINT FITTING FOR OTHER INTERACTION POWERS}
\label{sec:A4}
\begin{figure*}[tb]
\includegraphics[width=0.96\textwidth]{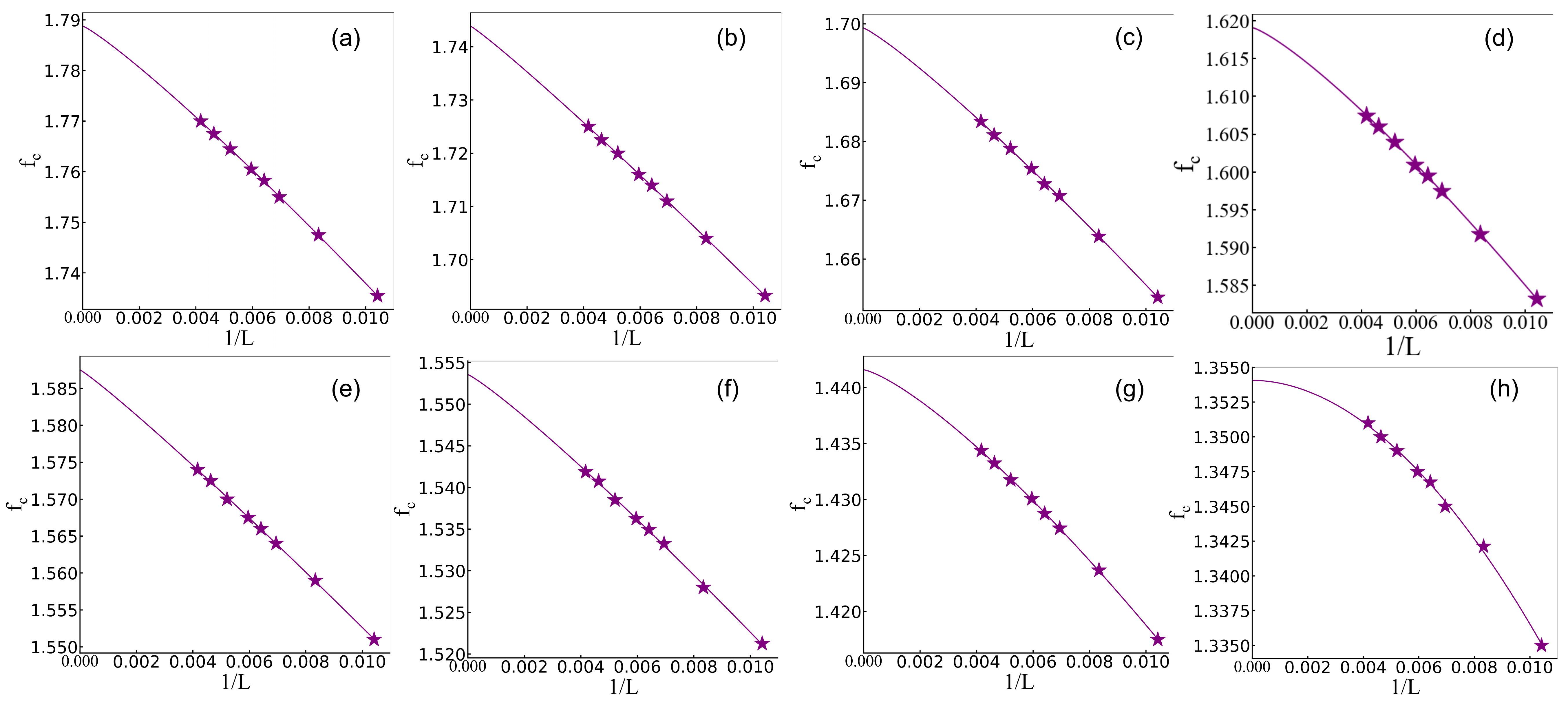}
\caption{(Color online) The finite size scaling of pseudo-critical point $f_{c}(L)$ as a function of  inverse system sizes $1/L$ for (a) $\alpha=1.3$,(b) $\alpha=1.35$,(c) $\alpha=1.4$, (d) $\alpha=1.5$, (e) $\alpha=1.55$, (f) $\alpha=1.6$, (g) $\alpha=1.8$, (h) $\alpha=2.0$. We use polynomial fitting formula: $\chi_{L}(f_{c}^{*})=L^{\mu}(a+bL^{-1})$}
\label{fig:appD}
\end{figure*}

In this section, we provide additional data to extrapolate accuracy critical points for other long-range interaction powers. 

\begin{figure*}[tb]
\includegraphics[width=0.96\textwidth]{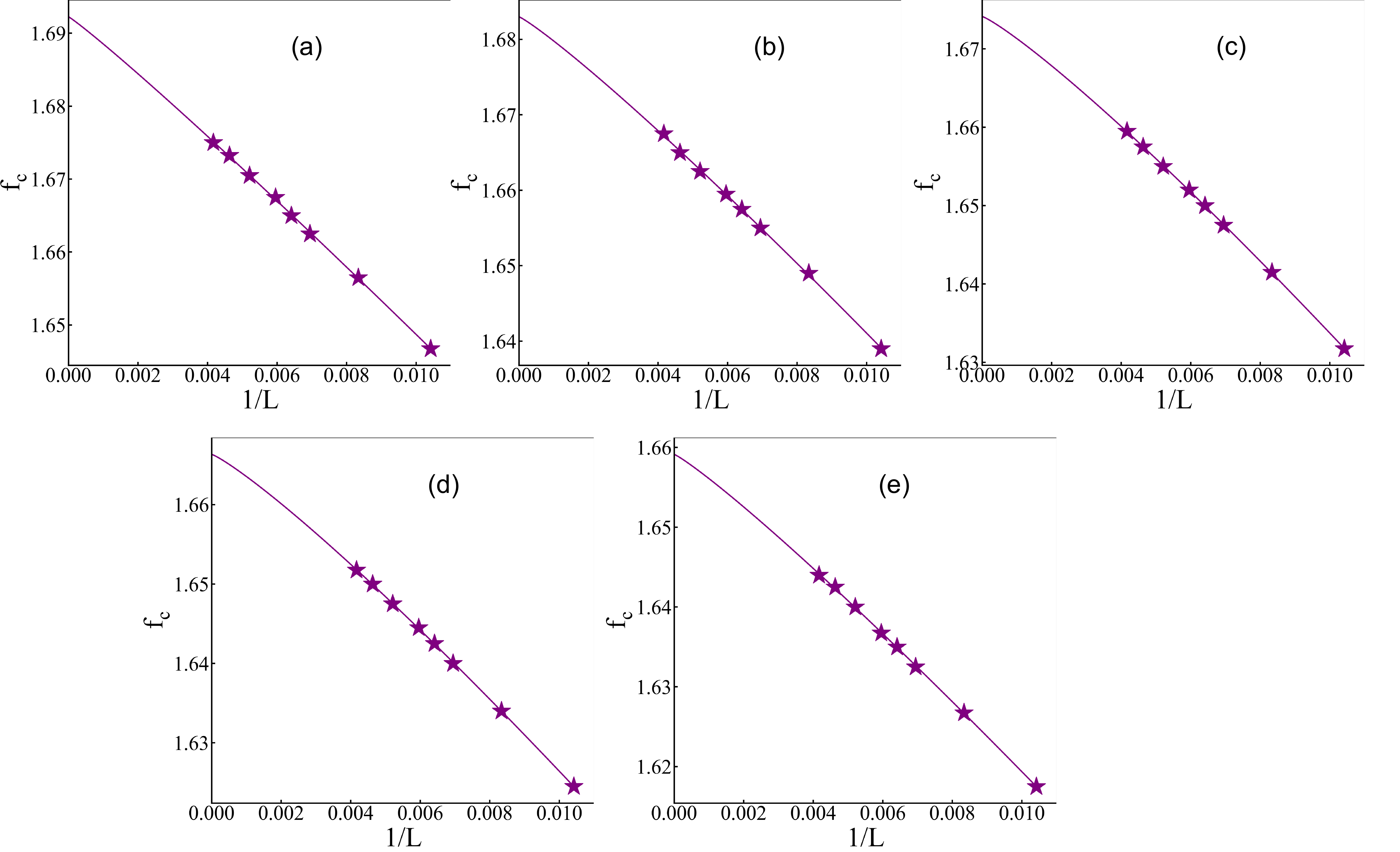}
\caption{(Color online) The finite size scaling of pseudo-critical point $f_{c}(L)$ as a function of  inverse system sizes $1/L$ for (a) $\alpha=1.41$,(b) $\alpha=1.42$,(c) $\alpha=1.43$, (d) $\alpha=1.44$, (e) $\alpha=1.45$. We use polynomial fitting formula: $\chi_{L}(f_{c}^{*})=L^{\mu}(a+bL^{-1})$}
\label{fig:appD2}
\end{figure*}

As the same in the main text, on the one hand, the finite-size scaling of pseudo-critical point $f_{c}(L)$ as a function of inverse system sizes $1/L$ for $\alpha=1.3$ (a), $\alpha=1.35$ (b), $\alpha=1.4$ (c),  $\alpha=1.5$ (d), $\alpha=1.55$ (e), $\alpha=1.6$ (f), $\alpha=1.8$ (g), $\alpha=2.0$ (h), and $L=96,120,144,156,168,192,216, 240$ sites, are shown in Fig.~\ref{fig:appD}. On the other hand, in order to determine critical $\alpha_c$, we also show the finite-size scaling of pseudo-critical point as a function of inverse system sizes for $\alpha = 1.41$ (a), $\alpha=1.42$ (b), $\alpha=1.43$ (c), $\alpha=1.44$ (d),$\alpha=1.45$ (e) in the Fig.~\ref{fig:appD2}.The extrapolated critical points are summarized in Table.~\ref{tab:exponents}. We clearly see that the critical point $f_{c}^{*}$ shifts to weaker values with increasing $\alpha$.

\end{appendix}



\end{document}